\documentclass[journal,10pt,letterpaper,english,5p]{elsarticle}
\usepackage[utf8]{inputenc}
\usepackage{amsfonts}
\usepackage{amsmath}
\usepackage{amssymb}
\usepackage{listings}
\usepackage{rotating}
\usepackage{color}
\usepackage{url}
\usepackage{subfigure}
\usepackage{balance}
\biboptions{sort&compress}
\usepackage[colorlinks,bookmarksopen,bookmarksnumbered,citecolor=red,urlcolor=red]{hyperref}

\lstdefinelanguage{JavaScript}{
  keywords={typeof, new, true, false, catch, function, return, null, catch, switch, var, if, in, while, do, else, case, break},
  keywordstyle=\bfseries,
  ndkeywords={class, export, boolean, throw, implements, import, this},
  ndkeywordstyle=\color{darkgray}\bfseries,
  identifierstyle=\color{black},
  sensitive=false,
  comment=[l]{//},
  morecomment=[s]{/*}{*/},
  commentstyle=\ttfamily,
  stringstyle=\ttfamily,
  morestring=[b]',
  morestring=[b]"
}

\begin{document}

\title{An Ant Colonization Routing Algorithm to Minimize Network Power Consumption}

\author[uvigo]{Miguel Rodríguez-Pérez\corref{cor1}}
\ead{miguel@det.uvigo.es}
 
\author[uvigo]{Sergio Herrería-Alonso}
\ead{sha@det.uvigo.es}

\author[uvigo]{Manuel Fernández-Veiga}
\ead{mveiga@det.uvigo.es}

\author[uvigo]{Cándido López-García}
\ead{candido@det.uvigo.es}

\address[uvigo]{Dept. Telematics Engineering, University of Vigo, Spain.}

\cortext[cor1]{Corresponding author. Address Telematics Engineering Dept.,
     Univ. of Vigo, 36310 Vigo, Spain. Tel.:+34~986813459;
     fax:+34~986812116.}

\begin{abstract}
  Rising energy consumption of IT infrastructure concerns have spurred the
  development of more power efficient networking equipment and algorithms.
  When \emph{old} equipment just drew an almost constant amount of power
  regardless of the traffic load, there were some efforts to minimize the
  total energy usage by modifying routing decisions to aggregate traffic in a
  minimal set of links, creating the opportunity to power off some unused
  equipment during low traffic periods. New equipment, with power profile
  functions depending on the offered load, presents new challenges for optimal
  routing. The goal now is not just to power some links down, but to aggregate
  and/or spread the traffic so that devices operate in their sweet spot in
  regards to network usage. In this paper we present an algorithm that, making
  use of the ant colonization algorithm, computes, in a decentralized manner,
  the routing tables so as to minimize global energy consumption. Moreover,
  the resulting algorithm is also able to track changes in the offered load
  and react to them in real time.

  \vspace{1ex}
  \copyright{} 2015 Elsevier Ltd. This manuscript version is made available under the \href{http://creativecommons.org/licenses/by-nc-nd/4.0}{CC-BY-NC-ND 4.0 license}\\DOI: \href{http://dx.doi.org/10.1016/j.jnca.2015.08.011}{10.1016/j.jnca.2015.08.011}.
\end{abstract}


\maketitle

\section{Introduction}
\label{sec:introduction}

The current Internet infrastructure draws far more power than needed
for its usual operation. At the same time, the network is still
growing, so this inefficiency translates to ever increasing power
demands with high monetary and environmental costs. For reference, the
overall energy consumption of all networking equipment just in the USA
in 2008 was estimated to be larger than
$18\,$TWh~\cite{lanzisera10:_data_networ_equip_energ_use} and the
estimated energy usage for the year 2020 in Europe is more than
$38\,$TWh~\cite{group08:_smart}.

These high energy demands have spurred successful research on all areas of
networking, from the link
level~\cite{reviriego09:_perf_eval_eee,herreria12:_gi_g_model_gb_energ_effic_ether,Sivaraman2014110,Khan2013965,Jung20143}
to the networking layer adapting the routing decisions, as suggested in
Gupta's seminal paper~\cite{gupta03:_green_of_inter}. However, these traffic
engineering proposals were initially constrained to the mere aggregation of
traffic during low activity periods to power off some devices, as that was the
only way a non-power aware device could be made to draw less power. From
there, many researches have followed this idea applying it to different
scenarios. Just as an example,
\cite{chiaraviglio12:_minim_isp_networ_energ_cost,caria12:_how,botero12:_energ_effic_virtual_networ_embed,addis14:_energ_manag_throug_optim_routin}
study centralized algorithms to minimize the number of active network
resources to get significant power savings, assuming a simple on-off power
model of networking equipment. Decentralized algorithms, as extensions to the
ubiquitous OSPF protocol, were also explored
in~\cite{cianfrani10:_energ_savin_routin_algor_green_ospf_protocy,cianfrani12:_ospf_integ_routin_strat_qos}.

Fortunately, newly produced networking equipment is increasingly becoming more
power aware. For instance, old Ethernet interfaces drew a fixed amount of
power regardless of the actual load. Since the arrival of the IEEE~802.3az
standard~\cite{802.3az}, this is no longer the case as they can adapt their
power demands to the traffic load. Thus, it is unnecessary to turn them off
completely in order to save
power~\cite{herreria12:_gi_g_model_gb_energ_effic_ether}. This trend is not
only limited to Ethernet devices. It also appears in optical
networks~\cite{zhang11:_towar_energ_effic_epon_epon,rodriguez12:_improv_energ_effic_upstr_epon},
switching modules~\cite{bianco12:_power_savin_distr_multi_router_archit}, etc.
The result is that new networking equipment exhibits non-flat power
profiles~\cite{herreria12:_gi_g_model_gb_energ_effic_ether}, and thus presents
an opportunity to regulate the traffic offered to each device, either
spreading or concentrating it, to take advantage of the power profile of each
device. These new capabilities are explored for instance
in~\cite{cardona09:_energ_profil_aware_routin,garroppo11:_energ_aware_routin_based_energ_charac_devic,seoane11:_energ_energ_effic_ether}.
The idea is not to concentrate traffic in a few set of links and power off the
rest, but to find the optimum share of traffic that minimizes energy costs
according to the power profile of each device.

In this paper we present the first dynamic decentralized algorithm capable of
adapting routing decisions to minimize energy usage when networking equipment
has otherwise unrestricted power profiles. Although it is not the first
proposal to make use of the ant colony optimization
algorithm~\cite{di97:_antnet} for energy saving~\cite{kim12:_ant_inter}, it is
the first that does not limit its routing decisions to decide the set of links
to power off for a given traffic matrix. In fact, it takes advantage of links
with non-flat power profiles and adjusts their traffic load in real time to
minimize power consumption while keeping all the installed capacity available.
This lets the network react better to unexpected spikes in the traffic load
and, additionally, improves the network resilience in case of a link failure.
The main difficulty in the adaptation of the original ant colony optimization
algorithm comes from the fact that, for the problem at hand, the cost of a
given route is not a simple linear function of its load and thus the protocol
becomes more complex than in the original version of the
algorithm~\cite{di97:_antnet}. We show in the next sections how this problem
was solved.

The rest of the paper is organized as follows. In
Section~\ref{sec:related-work} we present the related work. Then,
Section~\ref{sec:problem-statement} defines the problem in detail. Our
algorithm is described in Section~\ref{sec:trancas-algorithm}. Then,
an evaluation is carried out in
Section~\ref{sec:performance-evaluation} to finally present our
conclusions in Section~\ref{sec:conclusions}.

\section{Related Work}
\label{sec:related-work}

Research on new routing procedures that save power on communication networks
have been ongoing for a few years already. The first proposals focused on
concentrating the traffic on a reduced set of network elements so that unused
resources could be powered off during low load periods decreasing power
consumption. \cite{cianfrani12:_ospf_integ_routin_strat_qos} belongs to this
first family of proposals. It tries to concentrate traffic flows on a reduced
set of links to power off the rest. Another proposals in the same vein
are~\cite{chiaraviglio12:_minim_isp_networ_energ_cost} and~\cite{Yang20141}.
The first formulates a minimization problem of the energy consumption
considering that powered nodes and links need a constant amount of power, and
the second treats the problem of maximizing the number of powered off links.
As both problems are intractable
(NP-complete)~\cite{cianfrani12:_ospf_integ_routin_strat_qos,chiaraviglio12:_minim_isp_networ_energ_cost,kim12:_ant_inter,Yang20141},
both articles provide some heuristics to approximate the solution. All these
proposals, however, do not take into account the different power profiles that
new power-aware networking equipment exhibits and may even cause more harm
than good when these profiles are super-linear, as the increased power
consumption caused by traffic aggregation can surpass any power savings
obtained by the reduced consumption of the powered-down resources.

New proposals that take into account the different power profiles are also
known in the literature. For
instance,~\cite{chiaraviglioss:_model_sleep_mode_gains_energ_aware_networ}
considers super-linear energy costs functions in the analysis of the maximum
power savings attainable by powering down part of the network.
In~\cite{seoane11:_energ_energ_effic_ether} the authors formulate a
minimization problem considering the links formed by IEEE~802.3az links.
Similarly, the authors of~\cite{cardona09:_energ_profil_aware_routin} address
a similar problem and compare the results obtained with both super and
sub-linear power profiles. The same problem is also studied
in~\cite{garroppo11:_energ_aware_routin_based_energ_charac_devic,garroppo13:_does_traff_consol_alway_lead},
this time considering bundle links between adjacent routers. The authors find
out in~\cite{garroppo13:_does_traff_consol_alway_lead} that traffic
consolidation does not always lead to energy savings.

The main practical issue with all of these proposals is the complexity of the
problem, which is NP-complete~\cite{kim12:_ant_inter}. Finding the optimum
solution in a real network is very hard and it cannot be usually solved in a
short enough amount of time. NP-complete problems can be tackled employing
search heuristics, usually inspired by elements of the nature, that trade some
optimality in the found solution for execution time. In fact, such heuristics
have already been used with success in other areas of
networking~\cite{Nazi2014246}. So, there is a new line of research that
applies search heuristics to the route optimization problem reducing its
computational complexity. For
instance,~\cite{lu13:_genet_algor_energ_effic_qos_multic_routin} presents an
algorithm to save power in a restricted scenario of a multicast transmission
using genetic algorithms to find the solution to the routing problem.
In~\cite{galan2013:_using} the authors use the particle swarm optimization
technique to study the trade-off between the number of power profiles in line
cards and the energy savings realized. Finally,~\cite{kim12:_ant_inter} uses
the ant colony optimization algorithm to choose which links to power off to
maximize energy savings during low usage periods. Regretfully, none of these
works takes advantage of the energy savings capabilities present in current
equipment, unlike our proposal that permits the links to stay up, but
modulates their offered load to minimize energy consumption without affecting
the network resiliency.

\section{Problem Statement}
\label{sec:problem-statement}

We model the network as a directed graph $G = (N, \Lambda)$, with $N$ being
the set of nodes (i.e. IP routers) and $\Lambda$ the set of directed links.
Each link $\ell = (u, v) \in \Lambda,\,u,v \in N,$ with a nominal capacity
$\mu_{\ell},$ has associated a dynamic cost function $c_{\ell}(\rho_{\ell})
\in \mathbb{R}^+,$ with $\rho_{\ell}$ being the normalized traffic load
carried by the link. That is $\rho_{\ell} \triangleq \lambda_{\ell} /
\mu_{\ell}$, where $\lambda_{\ell}$ is the amount of traffic carried over the
link $\ell \in \Lambda$. Therefore, the cost of the links varies with the
offered load. Furthermore, we assume $c_{\ell}\left(\rho_{\ell}\right) =
\infty\text{ if } \rho_{\ell} > 1$.

The cost function captures the power needed to run the links at a given load.
Although most currently deployed links lack load-aware power profiles, new
links, such as those implementing IEEE~802.3az, have non-flat power profiles
that must be accounted for when implementing energy-aware routing protocols.
In our analysis we will assume that the power profile function is
monotonically increasing with link load. Also, for simplicity, the power
needed by the engines of the routers is assumed to be almost constant and so
it is absent from our analysis.

We will model the network traffic as a set of flows $\Phi$. Each flow $f \in
\Phi$ is described by a triple $f = \left(o, d, \lambda_f\right)$, with $o, d
\in N$ being the origin and the destination nodes respectively, and
$\lambda_f$ the amount of traffic carried by the flow. Each flow $f$ follows a
path $p_{\mathit{f}}$, defined as an ordered set of adjacent links going from
the origin node $o$ to the destination node $d$. There is a list of
symbols used in this article in Table~\ref{tab:symbols_legend}.
\begin{table*}
  \centering
  \begin{tabular}{c l}\hline
    \textbf{Symbol} & \textbf{Definition} \\\hline
    $N$ & Set of nodes in the network \\
    $E$ & Set of network edges\\
    $\Lambda$ & Set of links in the network \\
    $\mu_{\ell}$ & Nominal capacity of link $\ell$\\
    $c_{\ell}(\rho)$ & Cost function of link $\ell\in \Lambda$ for load $\rho$\\
    $\vec c^f$ & Direct costs of flow $f$\\
    $\vec \gamma^f$ & Indirect cost caused by flow $f$\\
    $\Phi$ & Set of flows \\
    $f(o, d, \lambda)$ & Flow from node $o$ to $d$ carrying traffic
    $\lambda$\\
    $p_{\mathit{f}}$ & Path followed by flow $f$\\
    $g_i^f(j)$ & \emph{Goodness} at node $i$ for taking $j$ as the next hop of
    flow $f$\\
    $\Gamma$ & Estimated network cost for the current agent\\
    $\pi_e$ & Threshold between random and \emph{goodness} based next node
    selection \\
  \end{tabular}
  \caption{Notation.}
  \label{tab:symbols_legend}
\end{table*}

The total cost of the network, that is, the amount of power needed to operate
it at any given time, can be computed as the sum of the costs of all the links
in the network. As the link cost function is not necessarily linear, each link
load must be obtained first. Let ${a(f, \ell)}$ be the route-link incidence
matrix, defined such that

\begin{equation}
  \label{eq:indices}
  a(f, \ell) \triangleq
  \begin{cases}
    1, & \mbox{ if } \ell \in p_{\mathit{f}}\\
    0, & \mbox{ otherwise.}
  \end{cases}
\end{equation}

Then, the load of a link is simply
\begin{equation}
  \label{eq:link-load}
  \rho_{\ell} = \frac{1}{\mu_{\ell}}\sum_{f \in \Phi} \lambda_f a\left(f, \ell\right),
\end{equation}

and the power cost of the whole network is
\begin{equation}
  \label{eq:network-cost}
  P = \sum_{\ell \in \Lambda} c_{\ell}\left(\rho_{\ell}\right).
\end{equation}

Formally, our goal is to solve
\begin{equation}
  \label{eq:min-goal}
  \min P = \min \sum_{\ell \in \Lambda} c_{\ell}\left(
    \frac{1}{\mu_{\ell}} \sum_{f \in \Phi} \lambda_f a(f, \ell)
    \right),
\end{equation}
that is, to minimize the overall power consumption $P$ subject to the usual
topological constraints:
\begin{subequations}
  \begin{multline}
    \label{eq:flow-conservation}
    \sum_{\ell = (i, j) \in \Lambda} a(f, \ell) - \sum_{\ell = (j, i) \in
      \Lambda} a(f, \ell) \\= 
    \begin{cases}
      1, & \quad \text{if $i = o_f$} \\
      -1, & \quad \text{if $j = d_f$} \\
      0, & \quad \text{otherwise.}
    \end{cases} \quad \forall f \in \Phi
  \end{multline}
  \begin{equation}
    \label{eq:resource-constraints}
    \sum_{f \in \Phi} \lambda_f a(f, \ell) \leq \mu_{\ell}, \quad \forall \ell \in \Lambda
  \end{equation}
  \begin{equation}
    \label{eq:variables}
    a(f, \ell) \in \{0, 1 \} \quad \forall f \in \Phi, l \in \Lambda
  \end{equation}
\end{subequations}
Equations~\eqref{eq:flow-conservation} are the flow-conservation constraints,
and~\eqref{eq:resource-constraints} are the physical constraints of the
network. The choice of integer values for the variables $a(f, \ell)$ means
that the flows are \emph{unsplittable}, i.e., each flow must follow a single
path through the network. 
\begin{figure*}
  \centering
  \hfill{}\subfigure[Optimum forwarding for logarithmic cost function.]{
    \includegraphics{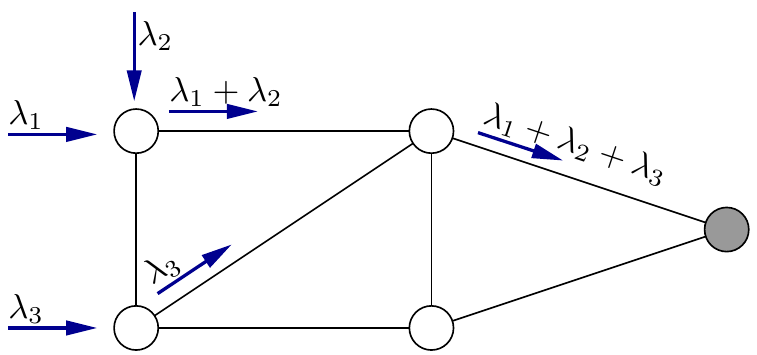}
    \label{fig:problem-statement-log}
  }\hfill{}
  \subfigure[Optimum forwarding for cubic cost function.]{
    \includegraphics{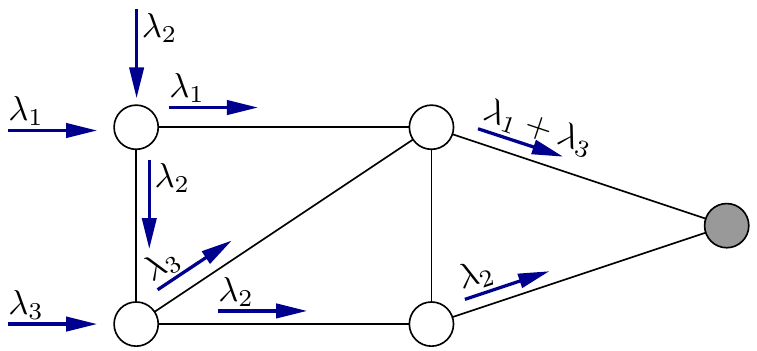}
    \label{fig:problem-statement-cubic}
  }\hfill{}

  \begin{tabular}{ll cc}\hline
    \textbf{Energy Profile}&\textbf{Cost Function}&\textbf{Total Cost for Figure~\ref{fig:problem-statement-log}}&\textbf{Total Cost for Figure~\ref{fig:problem-statement-cubic}}\\\hline
    Logarithmic&$c_{\ell}(\rho)=\log_{10}(1+\rho)$&\textbf{0.65}&0.85\\
    Cubic&$c_{\ell}(\rho)=\rho^3$&1.33&\textbf{0.48}\\\hline
  \end{tabular}  
  \caption{Power saving routing example for three flows
    $\lambda_1=\lambda_2=\lambda_3=1$ with a common destination (the gray
    node). Every link has the same cost function (logarithmic or cubic) and
    $\mu_{\ell}=3\, \forall \ell \in \Lambda$.}
  \label{fig:problem-statement-examples}
\end{figure*}
There is an example in Figure~\ref{fig:problem-statement-examples} that shows
feasible solutions to this problem for two different cost functions in a
simple five nodes network. This elementary example illustrates how the
different cost functions lead to distinct optimum forwarding strategies.

In the form above, the optimization problem is a generalization of the
\emph{unsplittable multicommodity flow problem} (UFP), where the
generalization consists in allowing \emph{arbitrary} cost functions
$c_{\ell}(\cdot)$ for the links. With linear costs,
\eqref{eq:min-goal}--\eqref{eq:variables} is the classical UFP, which in
specific instances is known to be NP-hard: for example, when the network $G$
has only one edge, the classical UFP specializes to the \textsc{knapsack}
problem.

For general link-cost functions, the relaxation of the
problem~\eqref{eq:min-goal}--\eqref{eq:variables} to $a(f, \ell) \in [0, 1]$,
namely to flows splitable over several routes, is generally a global
minimization problem. The case when $c_{\ell}(\cdot)$ are \emph{concave} functions
is, for instance, NP-hard~\cite{Sahni74,Varvalos90}. Since we do not impose
any prior assumption about the energy-consumption profiles, our problem can be
regarded NP from a computational perspective.

Our final goal is to design a new routing algorithm that
solves~\eqref{eq:min-goal}. The solution must be distributed and put low
requirements on the network nodes. Additionally, it must be able to
dynamically adapt to changing traffic demands and do so in a progressive
manner, such that the changes in the set of network paths do not lead the
network to a congested state nor cause undesired oscillations.

\section{Routing Algorithm}
\label{sec:trancas-algorithm}

As already stated in the introduction, we opt for a heuristic approach to
solve the aforementioned NP-hard optimization problem. Among all the families
of heuristic solvers available in the literature, the set of ant colony
algorithms~\cite{di97:_antnet} maps almost directly to the problem at hand.
Furthermore, their decentralized nature and time-adaptive characteristics are
requisites for any deployable solution. Note that we do not propose how the
routing decisions can be implemented in practice, although it is certainly
feasible on a MPLS~\cite{mpls-rfc} network where RSVP-TE~\cite{rsvp-te-rfc}
takes the job of setting up the label switched paths (LSPs).

As in the seminal AntNet algorithm~\cite{di97:_antnet}, our algorithm relies
on autonomous agents (\emph{ants}) that travel the network gathering enough
information to form optimal paths. Agents travel the network from source to
destination and back to the source. In their forward path, they explore
different routes to the destination to measure their costs. In their return
path, they update statistics at every node related to the fitness of the
next-hop node chosen in the previous forward way. The per-flow routing table
is finally calculated as the most appropriate next-hop for a given destination
at every node.

While the original AntNet algorithm used path delay as the cost function, we
use the power consumption. Moreover, AntNet only obtains a single path to a
destination from a given core node, while our algorithm must be able to
calculate different routes for every flow traversing each core node. This
complicates the problem as now the cost of the links does not depend on the
amount of traffic being carried by them in a linear way, and so it is
important to consider individual routes for every flow, even when they share a
common intermediate node and destination. Throughout the rest of this section,
we detail how agents work and what information they collect to obtain the set
of optimal paths for each flow that minimize global power consumption.

\subsection{Information Gathering}
\label{sec:inform-gath}

A key part of the algorithm relies on obtaining enough information about the
network state for updating the flow routes. It is the job of a \emph{forward
  agent} to gather this information with the help of the network nodes.

For this, a forward agent departs periodically from the source node $o$
towards the flow destination $d$. This agent carries information about the
current flow rate ($\lambda_f$) and the current flow path ($p_{\mathit{f}}$).
The agent walks the network towards the flow destination in a
non-deterministic manner to be detailed later. When the agent reaches a new
node, it records the identity of the node in an internal list of visited
nodes. At the same time, the agent calculates the marginal cost of carrying
the flow traffic across the link used in the last hop and stores it internally
as $\vec c^f[i]$,\footnote{We assume that agents use the memory in the visited
  node to store its state to then serialize it and transmit it to the next
  node as an IP packet.} with $i$ the index of the previous node. This depends
on the cost function of the link, the traffic already being carried by the
link and whether the link is part of current flow path. The exact procedure to
calculate the marginal cost is shown in Listing~\ref{lst:fwd_agent_comp} (see
function \texttt{calcCost}). Note that forward agents simply need that core
nodes maintain statistics about aggregate traffic load in their outgoing links
to calculate the marginal cost.

\begin{lstlisting}[float,label=lst:fwd_agent_comp,caption=Procedure for the
marginal cost calculation.]
calcCost = function($\ell$, $\lambda_f$, $p_f$, $\lambda_{\ell}$, $\mu_{\ell}$) {
    if ($\ell \in$ $p_f$) { /* link belongs to the current path of flow $f$ */
        /* Traffic in the link not from our flow */
	$\lambda_r$ = $\lambda_{\ell}$ - $\lambda_f$;
    } else {
        /* No traffic is from our flow */
        $\lambda_r$ = $\lambda_{\ell}$;
    }
    origCost = $c_{\ell}\left(\lambda_r/\mu_{\ell}\right)$;
    newCost = $c_{\ell}\left((\lambda_r + \lambda_f)/\mu_{\ell}\right)$;

    return newCost - origCost;
}
\end{lstlisting}

Before leaving the current node, forward agents need to decide which neighbor
node to visit next. There is a trade-off in this selection, because agents
should explore all the possible paths, but, at the same time, more resources
should be used to explore \emph{good} paths, where a \emph{good} path is the
one that the agent knows that demands less power than others. To achieve a
balance in this selection, forward agents use two procedures to select the
next visited node: one completely random using no previously obtained
information, and the other one based on costs calculated by other agents.
Which procedure to use is selected at random too. With some small probability
$\pi_e$ the agent chooses the first procedure ensuring that eventually all
possible routes are explored. With probability $1-\pi_e$ the next node is
chosen according to its \emph{goodness} relative to the flow~$f$. There is a
vector $\vec G_i(f)$ at every node $i$ that stores the goodness values for
each flow $f$ to all neighboring nodes. That is, $\vec G_i(f) = \{g_i^f(j)
\,\,\forall j \,| \, (i, j) \in \Lambda\}$. The goodness of each node is a
probability related to the estimated power consumption of the flow should it
select the neighbor node as part of its path. It must be stored in the nodes
where it is updated by the \emph{backward} agents.

Finally, if a cycle is detected after arriving to a new node, all the
information about the nodes visited and the links traveled since the previous
visit is deleted from the agent. A pseudo-code version of the algorithm
governing forward agents is provided in Listing~\ref{lst:fwd_agent}.

\begin{lstlisting}[float,label=lst:fwd_agent,caption=Forward agent
algorithm.]
for each $f(o, d, \lambda_f) \in \Phi$ {
    $F_{o \rightarrow d}$ = new ForwardAgent($f(o, d, \lambda_f)$, $p_f$);
    sendAgent($F_{o \rightarrow d}$);
}

sendAgent = function($F_{o \rightarrow d}$) {
    i = o;
    V = {}; // Set of visited nodes
    $\vec c$ = []; // Vector holding the link costs

    do {
	j = nextNode($G_i(f)$);
	$\ell$ = (i, j);
	if (i $\in$ V)
	    purgeCycle(V, $\vec c\,$);
	V = V $\cup$ i;
	/* $\hat \lambda_l$ is the estimated amount of traffic carried by link $\ell$ */	  
	$\vec c\,[\text{i}]$ = calcCost(l, $\lambda_f$, $p_f$, $\hat \lambda_l$, $\mu_{\ell}$);
	i = j; // Visit the next node
    } while (i != d);
}
\end{lstlisting}

\subsection{Information Dissemination}
\label{sec:inform-dissem}

The information gathered by the forward agent on its way to the destination
node (the marginal costs and the path actually traversed) is used to update
the goodness values at each of the intermediate nodes. The \emph{backward
  agent} is in charge of this update process while it travels back to the
origin node following exactly the reverse route recorded by the forward agent.

At an intermediate node $i \in N$ in the route of the backward agent from the
destination $d$ toward the origin $o$, the goodness value is updated based on
the cost of the partial path followed before by the forward agent from $i$ to
$d$. In turn, this cost is the sum of two components: a direct cost that
results from the addition of the flow's traffic to the downstream links from
$i$ on the path, and an indirect cost that measures the impact on the costs
that the remaining flows would see should the current flow $f$ departs
partially or totally from its current route.
\begin{enumerate}
\item The \emph{direct cost} is computed directly from the measurements taken
  by the forward agent as
  \begin{equation}
    \label{eq:cost_from_j}
    C^f_{\text{direct}}(i) = \sum_{k = i}^d \vec{c}^f[k],
  \end{equation}
  where $d$ is the flow destination and $\vec{c}^f[k]$ is the vector of
  measures recorded by the forward agent at node $k$.

\item When a flow leaves or changes its route, this might actually induce an
  increase in the marginal costs of other flows that were sharing the same
  links, particularly if the energy profile in those links is sub-linear. This
  possible increment is thus regarded as the \emph{indirect cost} of the
  partial route from $i$ to $d$ used by the forward agent. Specifically, the
  indirect cost is initialized at the destination $d$ when a backward agent is
  created, with value
  \begin{equation}
    \label{eq:extra_cost_init}
    C_{\text{indirect}}^f(d) = \sum_{\ell \in p_f} \vec{\gamma}^f[\ell]
  \end{equation}
  where
  \begin{equation}
    \label{eq:extra_cost_link}
    \vec{\gamma}^f[\ell] \triangleq \left( c_\ell(\lambda_\ell - \lambda_f) +
    c_\ell(\lambda_f) - c_\ell(\lambda_\ell) \right)^+
  \end{equation}
  is the sum of the marginal increases in energy consumption of all other
  flows traversing the link if the flow $f$ were to leave link $\ell$, and
  $\lambda_\ell$ is the total traffic carried by link $\ell$. Writing
  $\lambda_r \triangleq \lambda_\ell - \lambda_f$ for the remaining traffic
  that stays in the link after the departure of flow $f$, the cost
  change~\eqref{eq:extra_cost_link} is simply the difference between the cost
  due to the remaining traffic $c_\ell(\lambda_r)$ and the cost savings of
  shifting $\lambda_f$ units of traffic off its current operating point, i.e.,
  $c_\ell(\lambda_\ell) - c_\ell(\lambda_f)$. The value of
  $\vec{\gamma}^f[\ell]$ is not allowed to be negative, as would happen for
  links with super-linear (convex) cost functions,\footnote{Note that the
    argument of~\eqref{eq:extra_cost_link} is negative iff $c_\ell(\cdot)$ is
    a superadditive function.} in order to disincentivize that the flows
  change prematurely their paths. Such changes could lead to undesirable route
  flapping and network instability.

  The values of the vector $\vec{\gamma}^f$ are computed by the backward agent,
  for every visited link, as part of its reverse path. $\vec{\gamma}^f$ is
  stored at the source node of the flow, and it is carried by the forward
  agents to be used in the next round of the backward agents as follows: when
  the backward agent leaves node $j$ to visit node $i$, the indirect cost is
  updated
  \begin{equation}
    \label{eq:extra_cost_node_i}
    C^f_{\text{indirect}}(i) = \begin{cases}
      C^f_{\text{indirect}}(j) - \vec{\gamma}^f[(i, j)], \quad & \text{if $(i,
        j) \in p_f$} \\
      C^f_{\text{indirect}}(j), \qquad & \text{otherwise}.
    \end{cases}
  \end{equation}
  Therefore, the indirect cost decreases toward the source, and diminishes by
  an amount equal to the cost of leaving the links upstream from $i$ that flow
  $f$ really uses. As a result of this computation, the paths in which the
  departure of a flow $f$ would produce a larger cost to other concurrent
  flows are penalized in comparison to paths where this does not occur.
\end{enumerate}

The sum of the direct and indirect costs is the \emph{raw goodness} value
computed by the backward agent before leaving its current node, and stored
there
\begin{equation}
  \label{eq:goodness}
  \Gamma^f(i) = C^f_{\text{direct}}(i) + C^f_{\text{indirect}}(i).
\end{equation}
The goodness is used as a metric to select the best next-hop for every flow.
To this end, it is essential that paths with less energy demands yield
increasing goodness, but this condition is not guaranteed for the raw goodness
for a number of reasons. The raw values are noisy due to the measurement
process, and have to be normalized first for allowing the comparison with the
values computed by other backward agents for the same flow, possibly after
having explored different paths. Finally, some adjustment is needed to ensure
that the metric is monotonically decreasing in $\Gamma^f[i]$. In this paper,
we will apply the same mechanisms and problem-independent constants as [21] to
derive the routing metric.

To simplify notation, we set $\Gamma = \Gamma^f(i)$ for the rest of this
Section. First, $\Gamma$ is normalized by a scaled average of previous
measurements
\begin{equation}
  \label{eq:filter-scale}
  r^\prime = \min \{ \frac{\Gamma}{\alpha \overline{\Gamma}}, 1 \}
\end{equation}
where $\alpha > 1$ is a suitable attenuation constant and $\overline{\Gamma}$
denotes the average of past samples of $\Gamma$. Averaging smoothes the
measurements and reduces the variance, but this variance can still be high and
trigger instability in the routing decisions. Thus, $r^\prime$ is corrected
according to
\begin{equation}
  r^\prime_a = \begin{cases}
    r^\prime - \textrm{e}^{-\frac{a \sigma}{\overline{\Gamma}}} \quad &
    \text{if $r^\prime < 0.5$} \\
    r^\prime + \textrm{e}^{-\frac{a \sigma}{\overline{\Gamma}}} \quad & \text{otherwise}
  \end{cases}
\end{equation} 
where $\sigma$ stands for the standard deviation of $\Gamma$. This correction
is enforced only if the average $\overline{\Gamma}$ is considered reliable,
i.e., if the ratio $\sigma / \overline{\Gamma} < \epsilon \ll 1$. If the
average is not stable, $\sigma / \overline{\Gamma} \geq \epsilon$ then a
penalty factor is added
\begin{equation}
  r^\prime_a = \begin{cases}
    r^\prime +1 - \textrm{e}^{-\frac{b \sigma}{\overline{\Gamma}}} \quad &
    \text{if $r^\prime < 0.5$} \\
    r^\prime -1 - \textrm{e}^{-\frac{b \sigma}{\overline{\Gamma}}} \quad & \text{otherwise}
  \end{cases}
\end{equation}
where $b < a$. Finally, the metric $r^\prime_a$ is compressed via the power
law $r^\prime_a \leftarrow (r^\prime_a)^h$ and clipped to the interval $[0,
1]$.

After these nonlinear recalibration steps, node $i$ computes the
\emph{goodness routing metric} to each of its neighbors $j \in N$, $(i, j) \in
\Lambda$ as follows:
\begin{equation}
  \label{eq:final-goodness}
  g_i^f(j) \leftarrow g_i^f(j) + \begin{cases}
    (1 - r^\prime_a) \bigl( 1 - g_i^f(j) \bigr) \quad \text{agent comes
      from $j$} \\
    -(1 - r^\prime_a) g_i^f(j) \quad \text{otherwise}.
  \end{cases}
\end{equation}
It is easy to check that if $\sum_{j \in N, (i,j) \in \Lambda} g_i^f(j) = 1$
then the same condition holds after applying the update
rule~\eqref{eq:final-goodness}, so $g_i^f(j)$ can be conveniently interpreted
as the \emph{likelihood} of preferring neighbor $j$ as the next
hop.\footnote{Note that routing is deterministic, not random, and that the
  traffic of a flow is not split among several paths. Thus, even if $g_i^f(j)$
  is used as a probability by forward agents, all the traffic from a given
  flow chooses as the next hop the neighbor with the highest $g_i^f(j)$
  value.} A total goodness of $1$ is easily enforced by choosing as initial
values $g_i^f(j) = 1/n_i$ for every neighbor $j$ of node $i$, with $n_i$
adjacent nodes. Therefore, in absence of better a priori information,
initially each neighbor receives the same goodness as a credit.

\subsection{Obtaining the New Path}
\label{sec:getting-new-path}

Once the new goodness values have been calculated, the backward agent selects
the maximum as the next hop for the flow. This information is stored
internally by the agent.

When the agent finally reaches the source node for the flow, the information
about the next hops is employed to construct a new path. Because the backward
agent follows the reverse path of the forward agent, and the newly constructed
path is just the ordered collection of best next nodes, as determined by their
respective goodness values, for the visited nodes, this new path is not
necessarily connected. So before replacing the current path, the origin node
performs a connectivity test on it. For this, it can either rely on an
existing link state routing algorithm or send any kind of source routed probe
packet.

Even if the recorded path is dismissed, the work done by the agents is not
lost. Chances are high that a new forward agent eventually follows the best
path, as it is the one with the highest goodness values at every node, and
thus, the resulting backward agent will record the whole path.

\subsection{Memory Requirements}
\label{sec:memory-requirements}

It is important to characterize the memory requirements for storing all the
information related to the state of the agents and any auxiliary information
they may need. To this end, we detail the memory needs of edge nodes, regular
nodes and agents. Obviously, since agents are not physical entities, they
cannot really store any information, so they store it transiently in the nodes
they visit. In any case, we account for this separately from the memory needs
of regular nodes for clarity.

The forward agent carries the following information: a set of visited nodes,
the cost of the traveled links, the flow rate, the current path, and the extra
cost of leaving links in the current path. That is, information about the
current path and the traveled one. All this information depends solely on the
path lengths, and so it usually scales with the logarithm of the number of
nodes.

The backward agent does not carry much more information than the forward one.
It just stores the current path and, additionally, the extra cost values of
those links traveled that are part of the current path. Finally, it also holds
a copy of the path followed by the forward agent and it records the best next
hop node for the visited nodes. Again, all this information is proportional to
the path length, so it is independent of the number of flows.

The agents do use information stored in the nodes to communicate with other
agents and to obtain some basic information for their calculations. Source
nodes must store for every flow originating from them the current path of the
flow, its rate and the extra cost incurred when the flow leaves any of the
links currently traversed. The rate information scales linearly with the
number of flows departing from the node, while the path information and extra
costs scales with the product between the number of flows departing at the
node and the logarithm of the network size. We consider as a single flow all
traffic between a given pair of edge nodes, the total information stored at
the edges is still manageable. In the worst case, it is $|E|\log(N)$, with
$|E|$ the number of edge nodes.

Regular nodes, and edge nodes too, need to store additional information for
the agents to do their calculations. They need an estimate of the traffic
being sent across every outgoing link for the cost calculation. They also need
to store the information for the best node selection: the \emph{goodness}
vector and the cost statistics for each flow. Each goodness vector has an
entry for every outgoing link ($\frac{2|\Lambda|}{|N|}$ on average), so its
size should remain relatively small. However, the node must store a goodness
vector for every network flow. In the worst case, there can be as many as
$|E|(|E|-1)$ flows in the network, so this is clearly the limiting factor of
the algorithm. To lower this memory requirement, nodes could use some kind of
eviction policy to free memory associated to flows without recent activity.
All this information is summarized in Table~\ref{tab:memory-requirements}.
\begin{table}
  \centering
  \begin{tabular}{l c}\hline
    \textbf{Element}&\textbf{Needed storage}\\\hline
    Forward agent&$O(\log(N))$\\
    Backward agent&$O(\log(N))$\\
    Source node&$O(|E|\log(N))$\\
    Core node&$O\left(\frac{2|\Lambda|}{|N|}|E|^2\right)$\\\hline
  \end{tabular}
  \caption{Memory requirements.}
  \label{tab:memory-requirements}
\end{table}

\section{Evaluation}
\label{sec:performance-evaluation}

In this section we will analyze the performance of our algorithm. We start
with a set of simple experiments in a synthetic topology that highlights the
behavior of the algorithm for different cost functions. Then, we show the
results on more realistic network topologies.

All the results have been produced by an open source in-house simulator
available at~\cite{rodriguez13:_tranc}.\footnote{We refrained from writing a
  module for a general purpose network simulator as~\cite{ns-2} as the amount
  of new code would be on the same order.} Our simulator abstracts packet
level simulation details and considers the long time traffic averages known.
This speeds up the simulations while, at the same time, let us employ publicly
available traffic matrices that do not usually detail packet level details.
The simulator reads two configuration files: one describes the network
topology and link parameters and the second one controls the traffic
characteristics. The simulated links are described by their maximum traffic
capacity and their cost function. The cost function for a given link is
reduced for simplicity to the set of coefficients $\{a_0, \ldots, a_n\}$ in
the general formula
\begin{equation}
\label{eq:total_cost}
c_{\ell}(\lambda) = a_0 \log \lambda + \sum_{i=1}^{n}a_i \lambda^{i-1}.
\end{equation}
This formula lets us represent the main power profiles links are expected to
exhibit in the near
future~\cite{cardona09:_energ_profil_aware_routin,chiaraviglioss:_model_sleep_mode_gains_energ_aware_networ}:
sub-linear, like those of IEEE~802.3az
links~\cite{herreria12:_gi_g_model_gb_energ_effic_ether}; linear (although
this is not expected to be found in links, it can be used to account for the
power costs of the switch matrix of the routers), a constant component,
although this does not have any effect on the routing decisions, and
super-linear. These latter profiles, like cubic ones, have been found in
Ethernet interfaces applying dynamic voltage and frequency
scaling~\cite{Zhai:2004:TPL:996566.996798}. Finally, we do not consider an
on-off power profile as we need all links to be active to be able to send and
receive agents through them. In any case, with a suitable scaling factor, the
logarithmic profile can be made similar to the on-off profile. For the
algorithm configuration parameters, we used the constants provided
in~\cite{di97:_antnet}: $\epsilon=0.25$, $a=10$, $b=9$ and $h=0.04,$ that are
problem independent. In any case, it has been found that Ant Colonization
algorithms are quite resilient to changes in the configuration
parameters~\cite{dhillon07:_perfor_analy_antnet_algor}, so further tuning has
not been deemed necessary.

Given the inherent random behavior of the algorithm, each simulation has been
repeated 100 times, modifying the initial seed of the random generator of the
simulator. All the provided results show the averaged measure of a given
metric along with its 95$\,$\% confidence interval, except in those cases
where the interval was too small.

\subsection{Algorithm Behavior}
\label{sec:trancas-behavior}

The first set of results shows the behavior of the algorithm in a
regular network. The topology consists of a simple switching matrix of
$n$ steps connecting $n$ traffic sources to $n$ destinations. Every
link has the same cost function and unlimited capacity. The goal is to
check the results obtained by the algorithm in an otherwise
unrestricted scenario. Traffic consists of $n$ identical flows going
from each source to every destination, for a total of $n^2$ flows in
the network.

\begin{figure}
  \centering
  \includegraphics[width=\columnwidth]{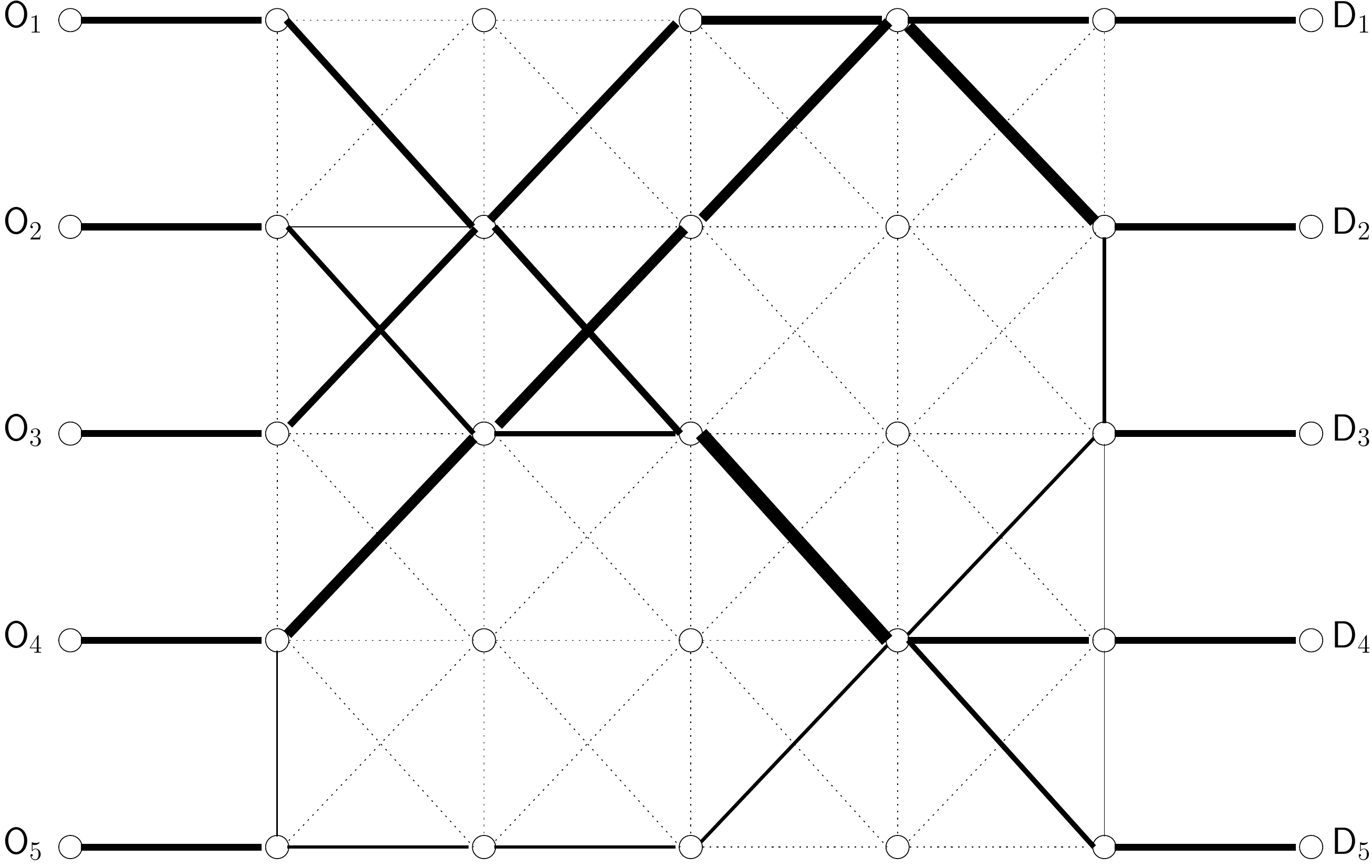}
  \caption{Algorithm behavior in a lattice network with a logarithmic cost function. Line
    width represents the number of flows in a link, while dotted lines show
    unused links.}
  \label{fig:log-multi-lattice}
\end{figure}

\begin{figure}
  \centering
  \includegraphics[width=\columnwidth]{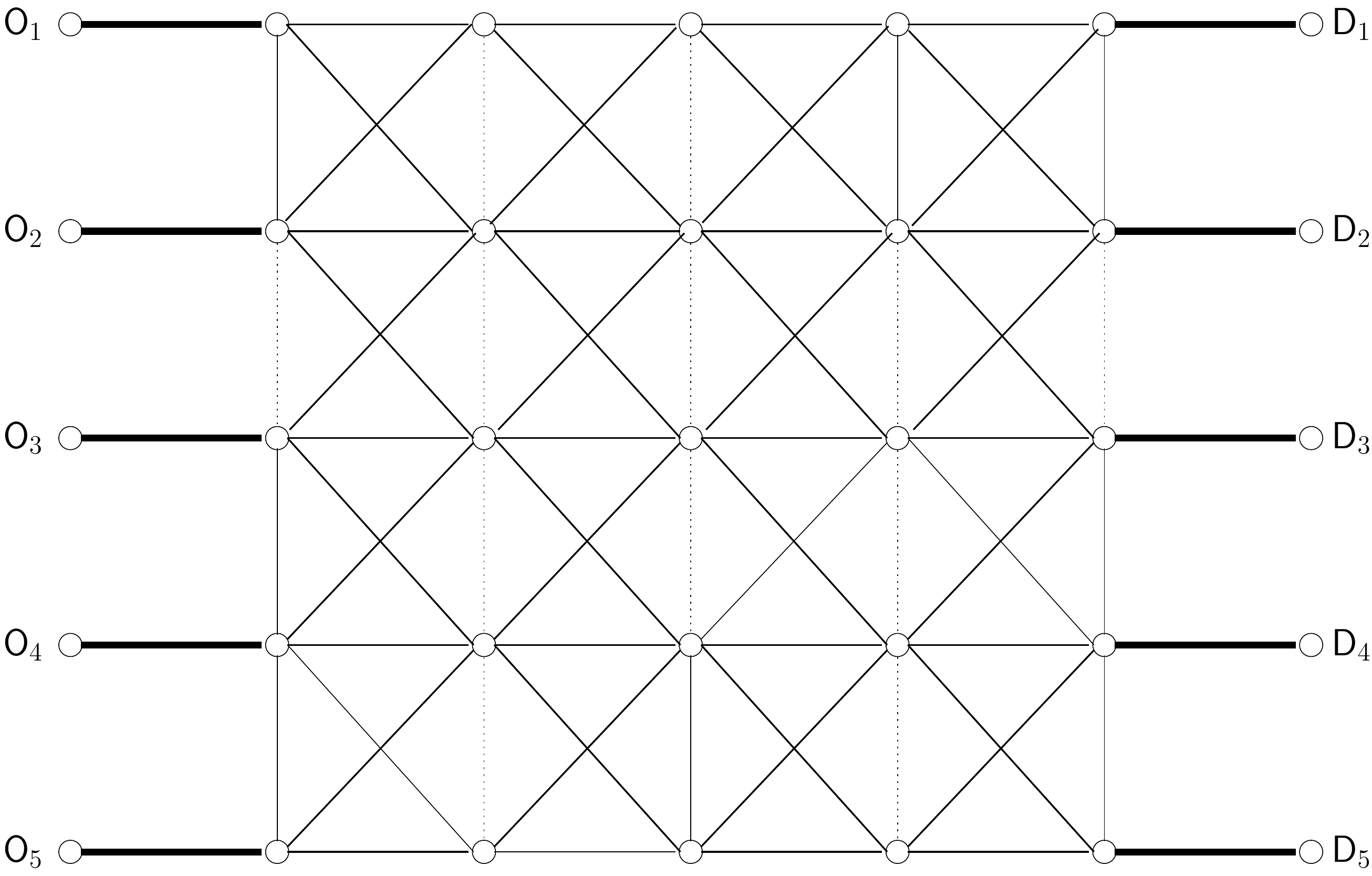}
  \caption{Algorithm behavior in a lattice network with a cubic cost function.}
  \label{fig:cubic-multi-lattice}
\end{figure}

\begin{figure}
  \centering
  \includegraphics[width=\columnwidth]{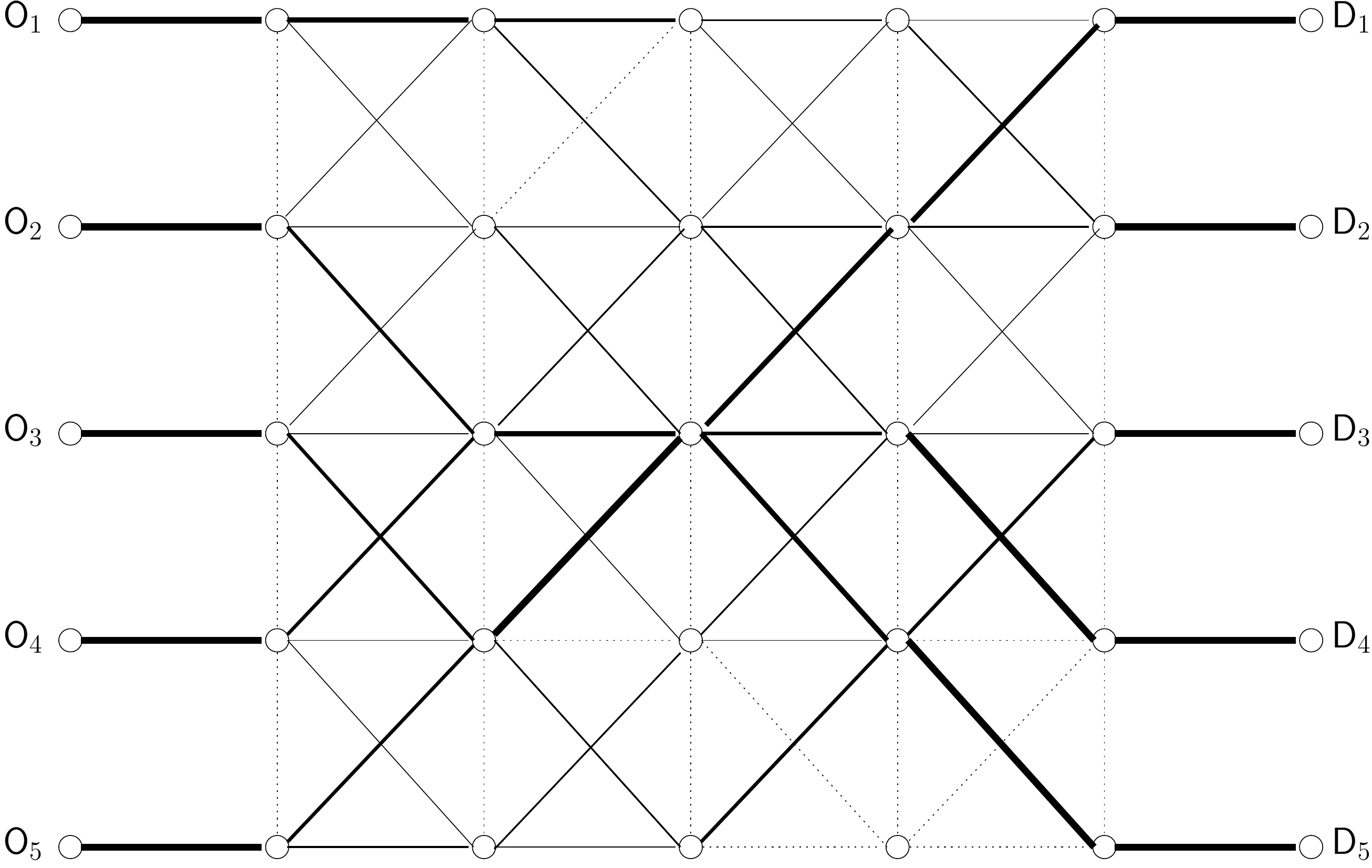}
  \caption{Algorithm behavior in a lattice network with a linear cost
    function.}
  \label{fig:linear-multi-lattice}
\end{figure}

Figures~\ref{fig:log-multi-lattice},~\ref{fig:cubic-multi-lattice}
and~\ref{fig:linear-multi-lattice} show a graphical representation of the
network and the link occupations for $n=5$ and three different cost functions:
logarithmic, cubic and linear, respectively. The number of flows in each link
is proportional to the line width, with dotted lines representing unused
links. After a close inspection of the graphics it can be seen that in the
logarithmic scenario (Figure~\ref{fig:log-multi-lattice}), routes tend to be
short (\emph{vertical} links are only used in the first and final steps) and
shared among various flows. Note that line widths are quite wide and, at the
same time, a lot of links remain unused. This is expected, as the marginal
cost of adding traffic to a link decreases with its load.

In contrast, for the cubic cost function
(Figure~\ref{fig:cubic-multi-lattice}), most links are lightly used. In fact,
almost all vertical links are employed to avoid sharing traffic on either
horizontal or diagonal links. Again, this is the expected behavior, as in this
case the marginal cost increases with load, so the algorithm must find the way
to spread the traffic across the net as long as the added cost (it employs
more links and longer routes) is not excessive.

In the linear scenario (Figure~\ref{fig:linear-multi-lattice}) the algorithm
just searches for the shortest routes regardless of how the links are shared
among flows. As in the logarithmic cost function scenario, there is almost no
single vertical link used, but it can be also observed that the load is not so
concentrated on a few links. In fact, the total number of empty links is
smaller.

We repeated these simulations on a somewhat larger net with $n=8$.
\begin{figure}
  \centering
  \subfigure[Logarithmic cost function]{
    \includegraphics[scale =1, width=\columnwidth]{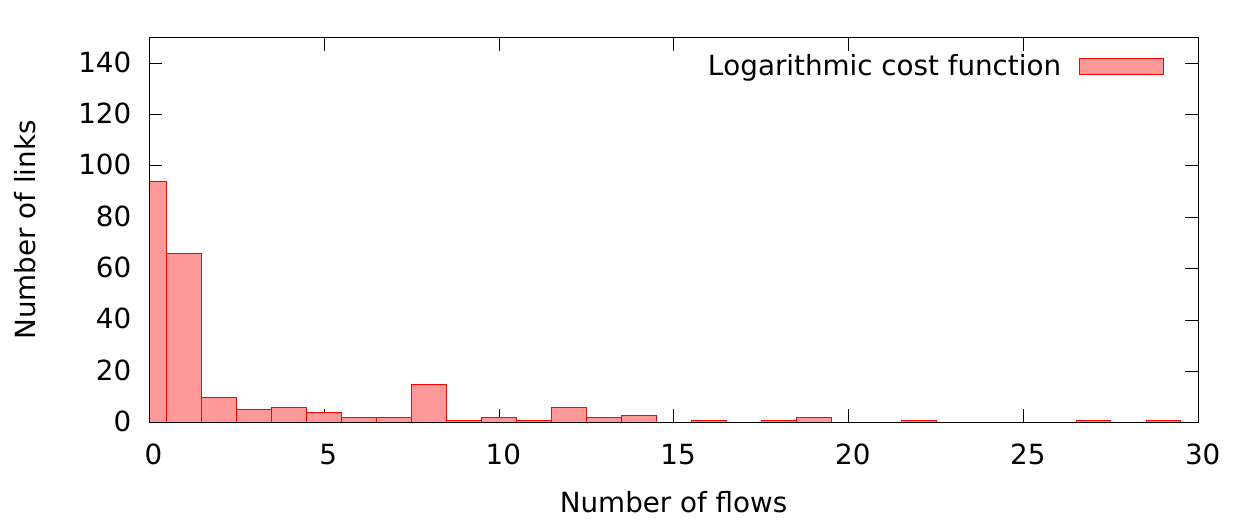}
    \label{fig:hist-lattice-8-log}
  }
  \subfigure[Cubic cost function]{
    \includegraphics[scale =1, width=\columnwidth]{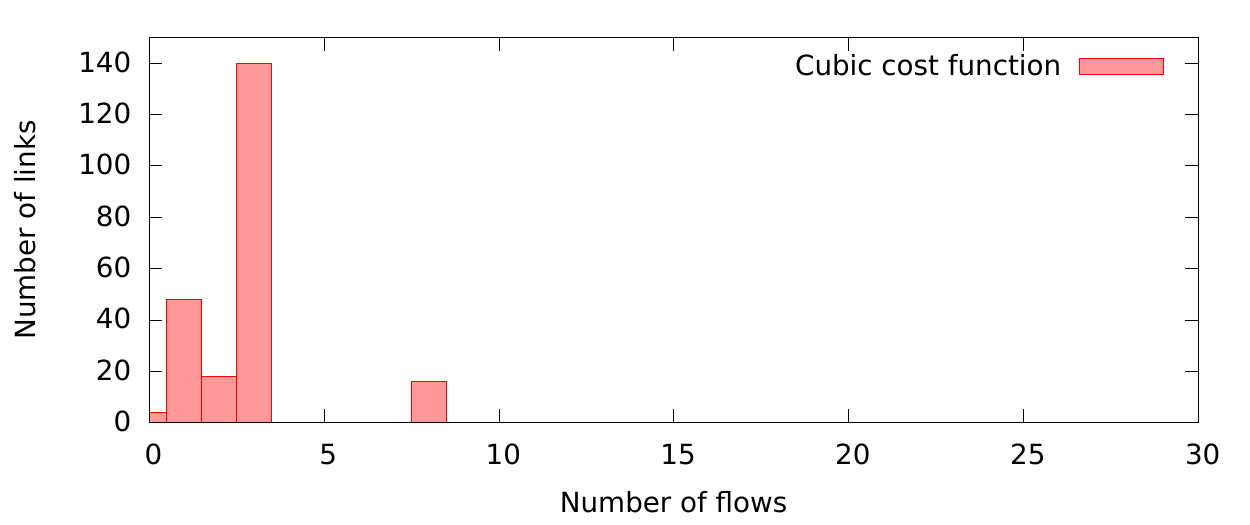}
    \label{fig:hist-lattice-8-cubic}
  }
  \subfigure[Linear cost function]{
    \includegraphics[scale =1, width=\columnwidth]{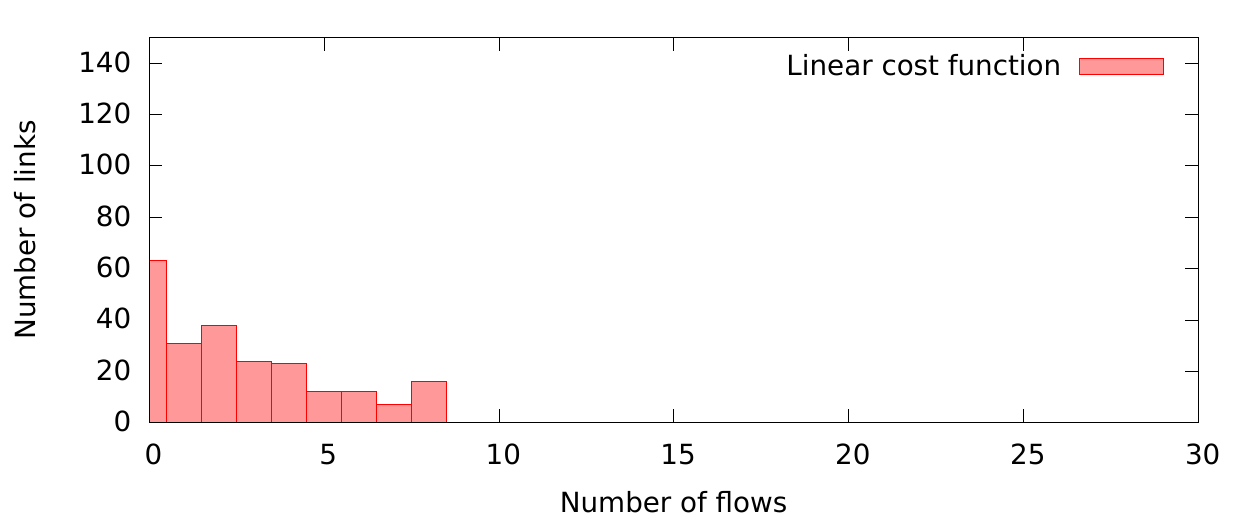}
    \label{fig:hist-lattice-8-linear}
  }
  \caption{Link occupation for an 8-nodes regular topology for different cost
    functions.}
  \label{fig:hist-lattice-8}
\end{figure}
Figure~\ref{fig:hist-lattice-8} shows how many links share a given number of
flows. Results agree with the above discussion. The logarithmic scenario has
the highest number of unused links (96) with some links carrying 27 or
even 29 flows. On the contrary, for the cubic cost function most links carry
just a few flows with almost no link sitting unused. The linear case, as
before, fits in between those scenarios.

\begin{table*}
  \centering  
    \begin{tabular}{|r|ccc|c|c|}
      \multicolumn{1}{}{}&\multicolumn{3}{c}{\textsc{Ant Colonization Algorithm}}&\multicolumn{1}{c}{\textsc{SPF}}\\\cline{2-5}
      \multicolumn{1}{r|}{Cost Function}&\textbf{Log}&\textbf{Linear}&\textbf{Cubic}&\textbf{Any}\\\hline
      \textbf{Path Length}&$9.9\pm0.6$&$9$&$9.9\pm0.1$&9\\
      \textbf{Energy Savings}&$13.3\pm1.3\,$\%&$0\,$\%&$69.9\pm0.1\,$\%&$0\,$\%\\\hline
    \end{tabular}
    \caption{Energy savings and path lengths obtained for different cost
      functions in a regular switching network with $n=8$. $95\,$\% confidence interval
      omitted for clarity when less than $0.1\,$\%.}
  \label{tab:lattice-8}
\end{table*}
Finally, the results obtained after the algorithm is run are summarized in
Table~\ref{tab:lattice-8}. It shows both the energy savings when compared to a
power unaware shortest-path-first (SPF) routing algorithm and the average
route lengths. As expected, for the linear cost function, the results are
identical to those of the SPF algorithm, and thus our algorithm produces no
energy savings, but keeps the optimum average path length of just nine hops.
However, for non-linear cost functions it pays a small penalty in path
lengths. This length increment is necessary to obtain more power efficient
routes. In fact, the energy savings for the cubic cost function ($69.9\pm0.1\,$\%) are
quite impressive in this topology.

\subsection{Performance Results}
\label{sec:performance-results}

We have also carried out experiments in more realistic network topologies, the
first set inspired in the topology of the old NSFNet network and a second one
in the \emph{nobel-eu} topology from the Survivable Network Design Library
(SNDlib)~\cite{SNDlib10}.

\begin{figure}
  \centering
  \includegraphics[width=\columnwidth]{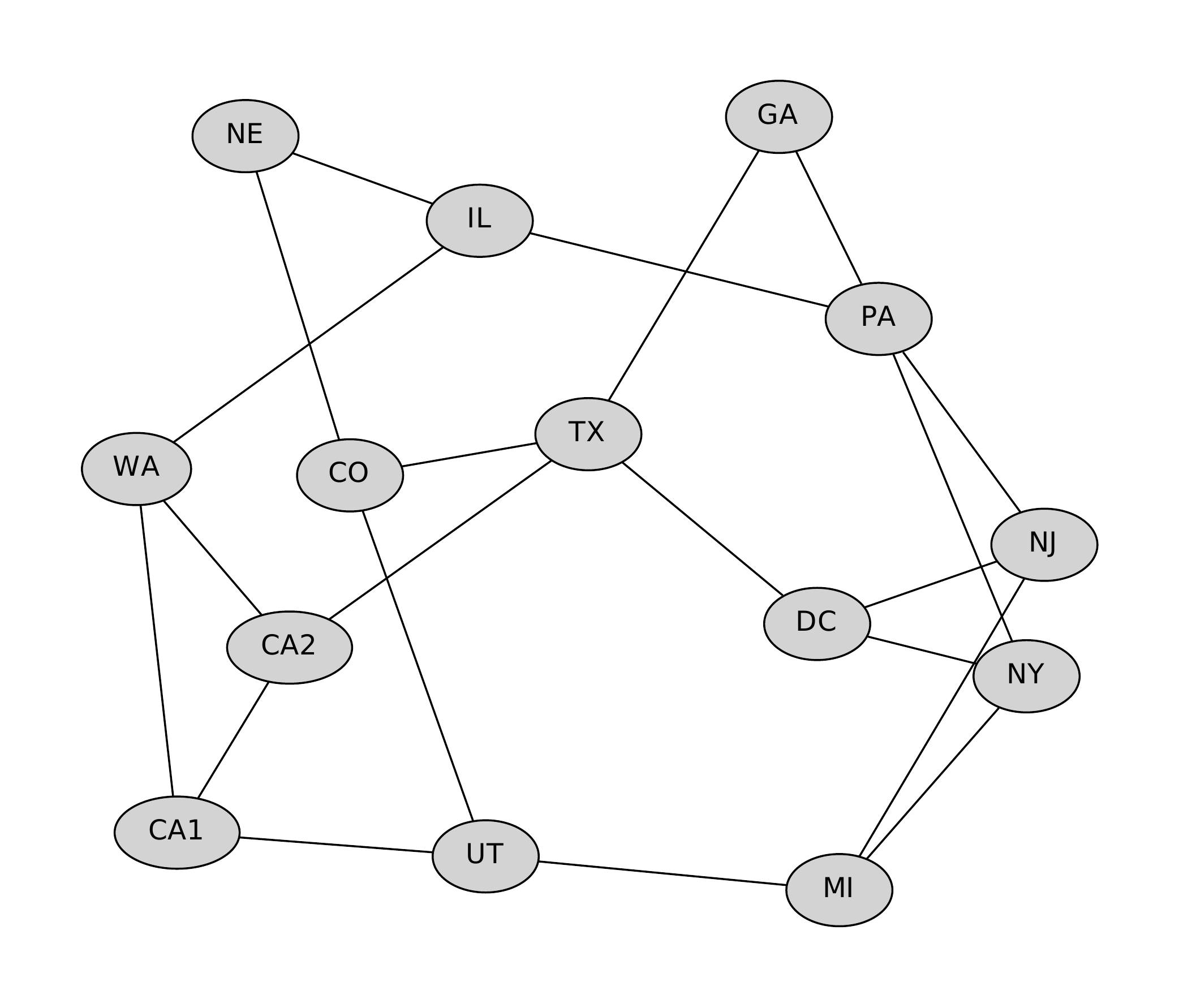}
  \caption{Network topology based on the original NSFNet. Node names
    correspond with their geographic location.}
  \label{fig:nsfnet}
\end{figure}

Figure~\ref{fig:nsfnet} shows the NSFNet network. We have conducted several
simulations with varying traffic matrices: a \emph{full-mesh} matrix with
traffic flowing from each source to every other destination; an
\emph{intra-coast} matrix, with traffic just between some nodes in the same
``coast''; and finally a \emph{coast-to-coast} matrix, with traffic flowing
from nodes in each coast to the other and vice-versa.
Although the traffic matrices can not be considered real by any means, they allow
to bring some light to the behavior and performance of the algorithm in a wide
range of representative scenarios.

\begin{figure}[t]
  \centering
  \includegraphics[width=\columnwidth]{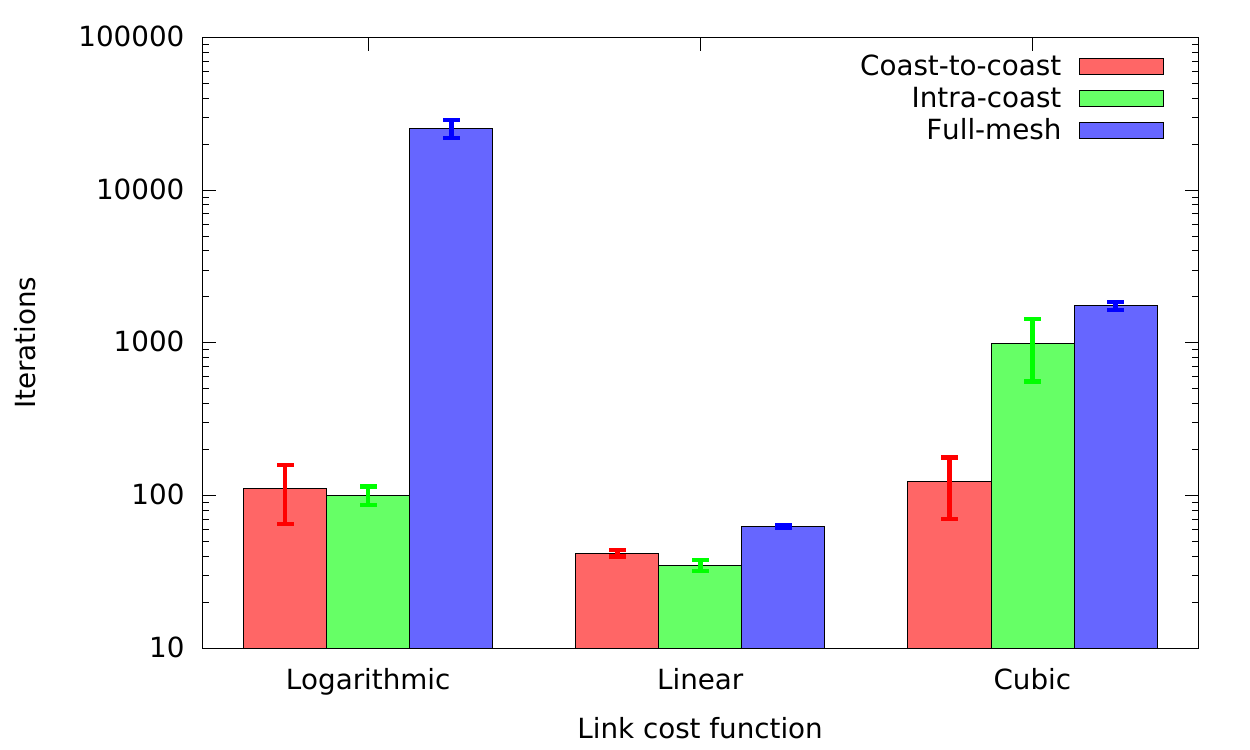}
  \caption{Number of iterations to reach 90\% of the greatest power savings for
    different traffic matrices and different link cost functions. Error bars
    show 95$\,$\% confidence intervals.}
  \label{fig:hist-it-nsfnet-110}
\end{figure}

\begin{figure}[t]
  \centering
  \includegraphics[width=\columnwidth]{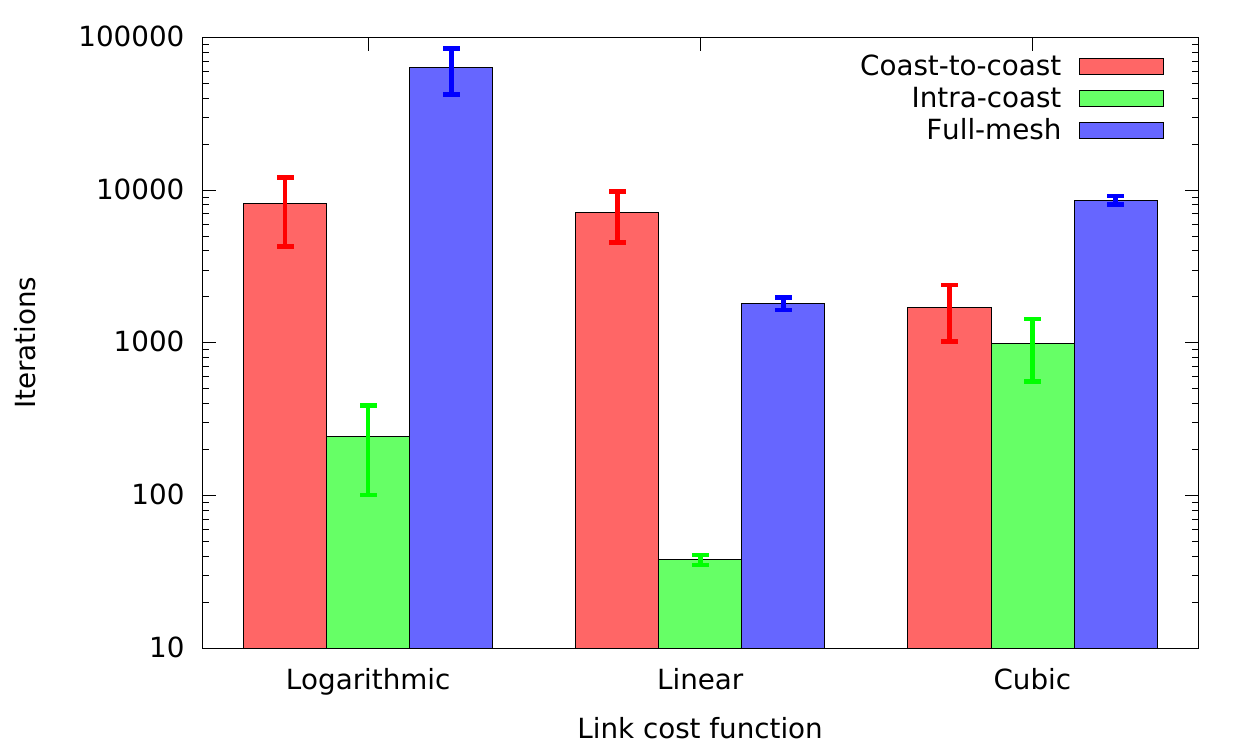}
  \caption{Number of iterations to reach 99\% of the greatest power savings for
    different traffic matrices and different link cost functions. Error bars
    show 95$\,$\% confidence intervals.}
  \label{fig:hist-it-nsfnet-101}
\end{figure}

The first performance characteristic we measured is the time needed by the
algorithm to reach $90\,$\% and $99\,$\% of the long term energy savings
it is able to achieve. We use the number of iterations, that is, the number of
forward agents sent by a source, as a proxy for this time, as it eventually
depends on the time separation between two consecutive agents. The results are
plotted in Figures~\ref{fig:hist-it-nsfnet-110}
and~\ref{fig:hist-it-nsfnet-101}. 

The first conclusion is that the algorithm is usually able to reach
the target of $90\,$\% quite fast. There is also a relationship
between the number of flows and the convergence speed. This can be
observed in the full mesh simulations, which usually take the largest
number of iterations. It can also be seen that the intra-coast
scenario is resolved very fast for any cost function, as it takes
almost the same number of iterations to reach $90\,$\% of the final
savings as to get to $99\,$\%. We believe that this is a consequence
of the optimal routes being quite short, and thus easy to come across
by the agents. On the other hand, both the coast-to-coast and the
full-mesh matrices need many more iterations to rise to the $99\,$\%
target.  For the linear cost function this happens because the optimal
routes are longer and thus the number of alternative routes with
similar costs is higher, lowering the likelihood of a forward agent to
follow them. The change for the logarithmic cost function is even
sharper. The reason is that not only the routes are longer, like in
the linear case. There is an additional complexity in the fact that
the algorithm tries to pack several flows in the same links for
maximum energy savings. As the agents take their routing decisions
autonomously it takes some extra time for routes to converge to the
same set of links. This also helps to explain why for the cubic cost
function the complexity increase is less noticeable. For super-linear
cost functions the greatest savings come from using disjoint routes,
so there is no need for several flows to converge on the same set of
links. So, it is easier for agents to choose links with low
occupation.

We have also measured two additional performance characteristics: actual power
savings and average path length increment. The power savings are compared to
the power consumed by a network using SPF as the power-agnostic routing
algorithm.
\begin{table*}
  \centering
  \begin{tabular}{c|l|rrr|}
    \multicolumn{2}{l}{}&
    \multicolumn{3}{c}{\textsc{Cost Function}}\\\cline{3-5}

    \multicolumn{2}{l|}{}&
    \multicolumn{1}{c}{\textbf{Log}}&
    \multicolumn{1}{c}{\textbf{Linear}}&
    \multicolumn{1}{c|}{\textbf{Cubic}}\\\cline{2-5}
    
    \textbf{Length} & \textsf{Coast to coast} & $20.1\pm1.1\,$\% & $0\,$\% &
    $16.1\pm1.2$\,\%\\
     \textbf{Increment}& \textsf{Intra coast} & $23.3\pm1.9\,$\% & $0\,$\% & $0\,$\%\\
    & \textsf{Full mesh} & $11.0\pm0.6\,$\% & $0\,$\% & $0.8\pm0.1\,$\%\\\cline{2-5}

    \textbf{Relative} & \textsf{Coast to coast} & $29.5\pm1.2\,$\% & $0\,$\% &
    $17.6\pm1.3\,$\%\\
    \textbf{energy} & \textsf{Intra coast} & $6.5\pm1.4\,$\% & $0\,$\% &
    $0\,$\% \\
   \textbf{savings}  & \textsf{Full mesh} & $29.9\pm0.5\,$\% & $0\,$\% &
   $12.8\pm0.06\,$\%\\\cline{2-5}
  \end{tabular}
  \caption{Performance improvement of the proposed algorithm for different traffic matrices in
    the network depicted in Figure~\ref{fig:nsfnet} when compared against
    Shortest Path First. $95\,$\% confidence
    intervals omitted for clarity when less than $0.1\,$\%.}
  \label{tab:nfsnet-summary}
\end{table*}
The results for the three traffic matrices and the three power profiles are
summarized in Table~\ref{tab:nfsnet-summary}. For the linear cost function,
the algorithm is unable to save more energy with regards to SPF, but this is
expected, as SPF discovers the optimal routes for these networks. In any case,
the results of our algorithm are also optimal, with no additional
energy demands nor increments in the path lengths.

In the logarithmic link cost networks, the algorithm obtains more than
$20\,$\% energy savings for the more complex traffic matrices. The route
lengths also grow, although the increments are below $25\,$\%. 

Finally, the cubic cost function does not attain any savings for the
intra-coast traffic matrix. This is because the shortest path routes
are already optimal. In fact, the path lengths are identical for both
the proposed algorithm and the SPF routing algorithm. For the rest of
the traffic matrices it gets savings in the $10$--$20\,$\% range by
distributing flows in different links, at the cost of an obvious
increment in the average path length.

In short, the proposed algorithm is able to trade some increment in route
lengths to save energy in the network. When the routes computed by a shortest
path first algorithm are already optimal, the routes computed by our proposal
are never worse: both average path length and energy consumption remain
identical.

As already stated at the beginning of the Section, we have also used a real
topology both to assess the behavior of our algorithm and to compare it
against the optimization shown
in~\cite{garroppo11:_energ_aware_routin_based_energ_charac_devic,garroppo13:_does_traff_consol_alway_lead}
and to those power profile unaware algorithms that minimize the number of
active
links~\cite{cianfrani12:_ospf_integ_routin_strat_qos,chiaraviglio12:_minim_isp_networ_energ_cost,kim12:_ant_inter,Yang20141}.
We have employed the topology and average traffic matrix of the
\emph{nobel-eu} core network from the SNDlib archive used in those works. The
\emph{nobel-eu} network is a European network consisting on 28 nodes connected
by 41 links and the traffic matrix consists on a total of 378 flows. For the
sake of the comparison, we have simplified the network model proposed
in~\cite{garroppo13:_does_traff_consol_alway_lead} as we restrict the number
of links between a given pair of nodes to one, albeit with unlimited capacity.

\begin{figure}
  \centering
  \includegraphics[width=\columnwidth]{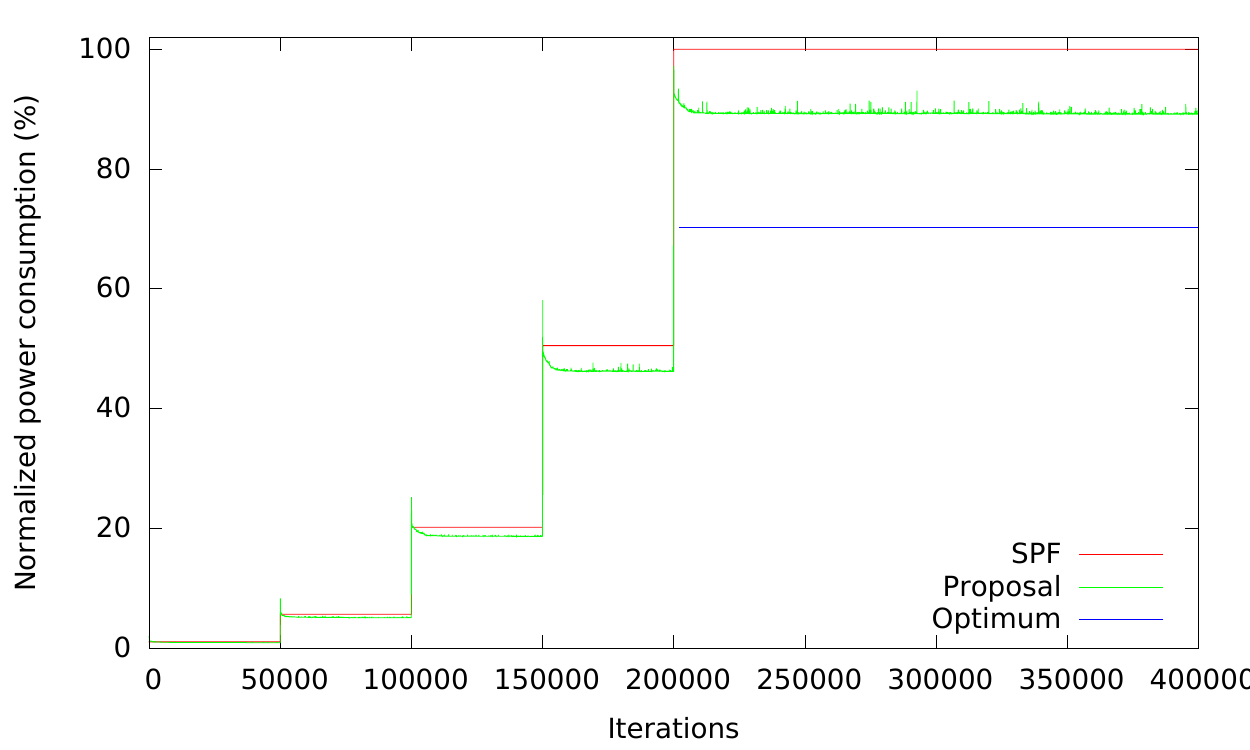}
  \caption{Comparison between our algorithm, SPF and the optimal result for
    the nobel-eu network with a cubic cost function.}
  \label{fig:nobeleu-cubic-evolution}
\end{figure}

Figure~\ref{fig:nobeleu-cubic-evolution} shows the normalized power consumption
with a cubic cost function.
Traffic was added to the network in several steps to show the dynamics of our
algorithm. It can be seen that the consumption raises briefly above that of
SPF when new traffic enters the network, but rapidly stabilizes below it after
a few iterations. For completeness we have calculated with the help of the IBM
CPLEX solver the optimum power consumption obtained considering our simplified
model of~\cite{garroppo13:_does_traff_consol_alway_lead}. As expected, the
centralized calculation is able to obtain the best results, albeit it cannot
adapt automatically to changing network conditions. We also performed the
experiment with a logarithmic cost function. In this case, our algorithm just
managed to save $3\pm1\,\%$ of the power needed when using SPF routes, while
the CPLEX solver managed to save 16\% of the power, considering again a static
scenario.

To simulate the results of the power profile unaware algorithms we eliminated
all but the most used outgoing link for each node when using SPF
routing.\footnote{The optimum result in these algorithms is obtained with
  unlimited capacity links, as a single outgoing link is enough to transmit
  all the traffic from a given node, and the rest of the links can be powered
  down.} Then, we calculated the global power usage in the modified graph. We
found that energy usage increases $9.5\,$\% for links with logarithmic cost
function when compared with the unmodified network using SPF as a routing
algorithm. This is due to the increased average length of the routes, resulting in
traffic consuming energy in more links. Results, however, are much worse for a
cubic cost function. In this case, traffic should be spread over various links
to minimize consumption, however, with a single path between each pair of
nodes this is not feasible. Energy consumption is $8.8\,$ times higher than in
the unmodified network. Although the results may seem counter intuitive, there
are to be expected, as all these algorithms are designed for networks with
fixed cost links.

\section{Conclusions}
\label{sec:conclusions}

In this paper we have presented a modified version of the
AntNet~\cite{di97:_antnet} algorithm to calculate, in a decentralized way,
optimal routes to reduce power consumption of network links. The presented
solution does not put any restriction in the power profile functions of the
networking equipment.

The proposal was tested in both synthetic and real scenarios with different
power profiles. The obtained results show power savings in the 10--20\% range
for real networks and up to 70\% in favorable, although unlikely, scenarios.
Moreover, the convergence times are small, as the 90\% of the savings are
usually obtained in less than 1000 iterations. Thus the algorithm can be used
continuously in background in the network, adapting the routing tables to the
medium-term averages of the traffic load of the incoming flows.

Finally, the results also show that it is necessary to take into account the
power profile of the links, as not doing so and blindly powering off less used
links can even augment power usage.

\section*{Acknowledgments}
\label{sec:acknowledgments}

Work supported by the European Regional Development Fund (ERDF) and the
Galician Regional Government under agreement for funding the Atlantic Research
Center for Information and Communication
Technologies (\href{http://atlanttic.uvigo.es/en/}{AtlantTIC}).


\bibliographystyle{elsarticle-num}
\bibliography{IEEEfull,trancas}

\end{document}